\documentclass[aps, apl, a4paper, reprint, superscriptaddress]{revtex4-2}
\usepackage{graphicx}
\graphicspath{{./Bilder/}}
\usepackage{mathtools}
\usepackage{amsfonts}
\usepackage{framed}
\usepackage{amssymb}
\usepackage[utf8]{inputenc}
\usepackage{subcaption}
\usepackage[all]{nowidow}
\begin{document}
\title{Thermostat for a relativistic gas}
\author{Noah Kubli}
\affiliation{Institut für Theoretische Physik, ETH Zürich}
\author{Hans J. Herrmann}
\affiliation{PMMH, ESPCI, 7 quai St. Bernard, 75005 Paris, France}
\affiliation{\mbox{Dept. de Física}, UFC, Fortaleza, Brazil}
\begin{abstract}
	\null
	\begin{center}
		\normalsize \emph {In memory of Dietrich Stauffer}
	\end{center}
	Molecular dynamics simulations of a three dimensional relativistic gas
	with a soft potential 
	are conducted with 
	different interactions and particle masses. 
	For all cases the velocity distribution agrees numerically with the Jüttner distribution.
	We show how the relativistic gas can be
	coupled to a thermostat to simulate the canonical ensemble at a given temperature $T$.
	The behaviour of the thermostat is investigated as a function of the
	thermal inertia and its appropriate range is determined
	by evaluating the kinetic energy fluctuations.
\end{abstract}
\maketitle

\newpage
\section{Introduction}
The velocity distribution of a gas is classically described 
by the Maxwell-Boltzmann distribution. However, this
distribution does not hold in special relativity as can be seen
from the fact that there would be a finite probability for particles
to exceed the speed of light. This problem becomes more severe 
for higher temperatures.

\begin{figure}[h]
	\includegraphics[width=0.3\textwidth]{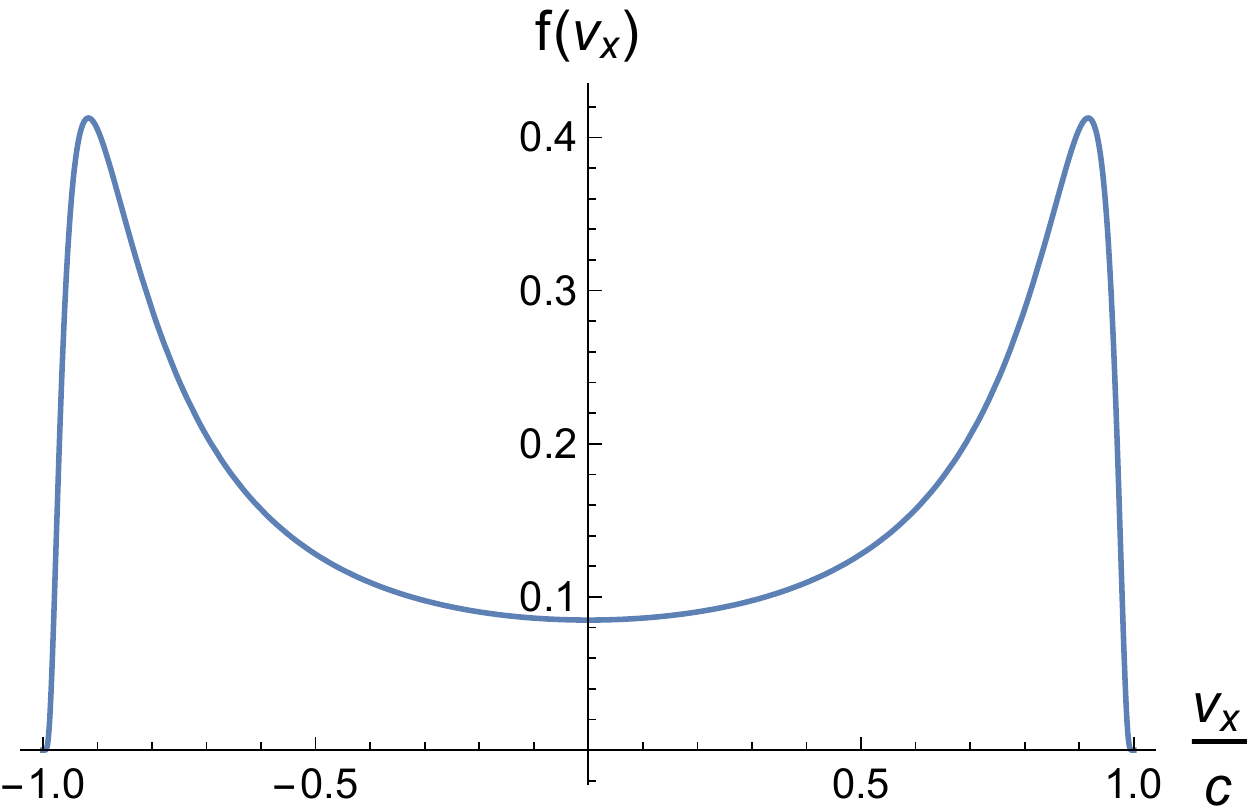}
\caption{\label{Jut2}
	Jüttner distribution in terms of one velocity coordinate
	$v_x$ (x-axis) in the ultrarelativistic regime ($k_{\rm B} T = 0.5$ and $m=1$).
	Two peaks arise as opposed to the one-peaked Maxwell-Boltzmann
	distribution which only becomes broader at high temperatures.
	}
\end{figure}

Jüttner generalised the Maxwell-Boltzmann distribution
in 1911 to relativistic gases \cite{juttner}.
Maximizing the entropy, he derived the following velocity distribution: 
\begin{equation}
	f(v_x, v_y, v_z) = \frac{1}{Z(m, k_{\rm B} T, N)} m^3 \gamma (v)^5 \exp \left(-\frac{m c^2 \gamma(v)}{k_{\rm B} T}\right)
\end{equation}
where $T$ is the temperature, $k_{\rm B}$ the Boltzmann constant, $m$ the mass of the particles,
$N$ the number of particles, $\gamma(v) = 1/\sqrt{1-v^2/c^2}$ the Lorentz factor and $c$ the
speed of light. $Z$ is a normalising factor.
The distribution is isotropic since it only depends on the
absolute value of the velocity $v$. It is straightforward
to obtain the distribution function of the absolute velocity $\lvert v\rvert$
through integration over the solid angle. One obtains:

\begin{equation}
	f(\lvert v\rvert) = \frac{4 \pi}{Z(m, k_{\rm B} T, N)} m^3 \gamma (v)^5 v^2 \exp \left(-\frac{m c^2 \gamma(v)}{k_{\rm B} T}\right)
\end{equation}
For low temperatures the Jüttner 
distribution in terms of $\lvert v\rvert$ resembles the Maxwell-Boltzmann distribution but for high 
temperatures a sharp peak right below the light speed arises
reflecting the fact that no particle can exceed the speed of light (see Fig. \ref{Jut2}).

In 2007 Cubero et al. conducted simulations in one dimension using collisional 
dynamics \cite{cubero}. Their simulation results agree 
very well with the Jüttner distribution and they could rule out 
another proposed covariant distribution introduced in Ref. \cite{dunkel}.
The same result was found using two dimensional collisional dynamics 
in a paper by Ghodrat and Montakhab \cite{montakhab}.
There a stochastic thermostat was used to simulate the canonical ensemble.

A problem concerning the Jüttner distribution is its non-Lorentz invariance:
It depends on the energy of the particle $mc^2 \gamma$ and on $\gamma ^5$
whereas a Lorentz invariant distribution would have to be 
dependent on $\gamma ^4$ \cite{evaldo}. To overcome
this problem, in Ref. \cite{evaldo} another covariant distribution
was proposed that depends on the rapidity instead of the velocities. 
The rapidity depends on the square of the velocities of the particles
relative to each other which is a Lorentz invariant quantity.

Applications of the Jüttner distribution can be found in astrophyics and cosmology,
for example in reconstructing the thermal history of  the universe \cite{astro}.
A recent paper investigated a special property of the Jüttner distribution
in solid state physics \cite{herrmann}:
At high temperatures the Jüttner distribution in terms of a single coordinate $v_x$
exhibits two peaks, one close to $c$ and one close to $-c$ (see fig. \ref{Jut2}). This
differs from the low temperature regime (and from the Maxwell-Boltzmann distribution)
which only broadens
when the temperature is increased. In Ref. \cite{herrmann} 
a critical temperature $k_{\rm B} T_c = (d+2)^{-1}$ (with $c=m=1$ and d the dimension)
is found at which the distribution changes from a single peaked
function to a double-peaked one. Since at the Dirac point of graphene electrons 
behave like a relativistic gas, effects 
of this transition could be measured in graphene when changing the 
Fermi energy \cite{herrmann}.

Here, we will introduce a deterministic relativistic thermostat to simulate
the canonical ensemble and verify its properties with
three-dimensional simulations using the relativistic equations of motion.

\subsection*{Temperature of a  relativistic gas}
For a classical ideal gas, the following well-known relation 
between temperature and kinetic energy holds: 

\begin{equation}
	\label{class_temp}
	k_{\rm B} T = \frac{2}{3N}E_{kin}
\end{equation}
with $N$ the particle number.
If we include interactions between the particles (i.e. a potential energy $V$)
$E_{kin}$ must be substituted by its time-average $\langle E_{kin} \rangle$ for the formula to hold.
To see what happens in the case of special relativity,
the equipartition theorem can be applied:
\begin{equation*}
	k_{\rm B} T= \langle {p_{i, \alpha}} \frac{\partial \mathcal{H}}{\partial p_{i, \alpha}} \rangle
\end{equation*}
where $\mathcal{H}$ is the Hamiltonian and 
$\vec{p}_i = (p_{i, \alpha})_{\alpha \in \{x, y, z\}}$
the momentum of particle $i$.

\begin{figure}[h]
\centering
\includegraphics[width=0.4\textwidth]{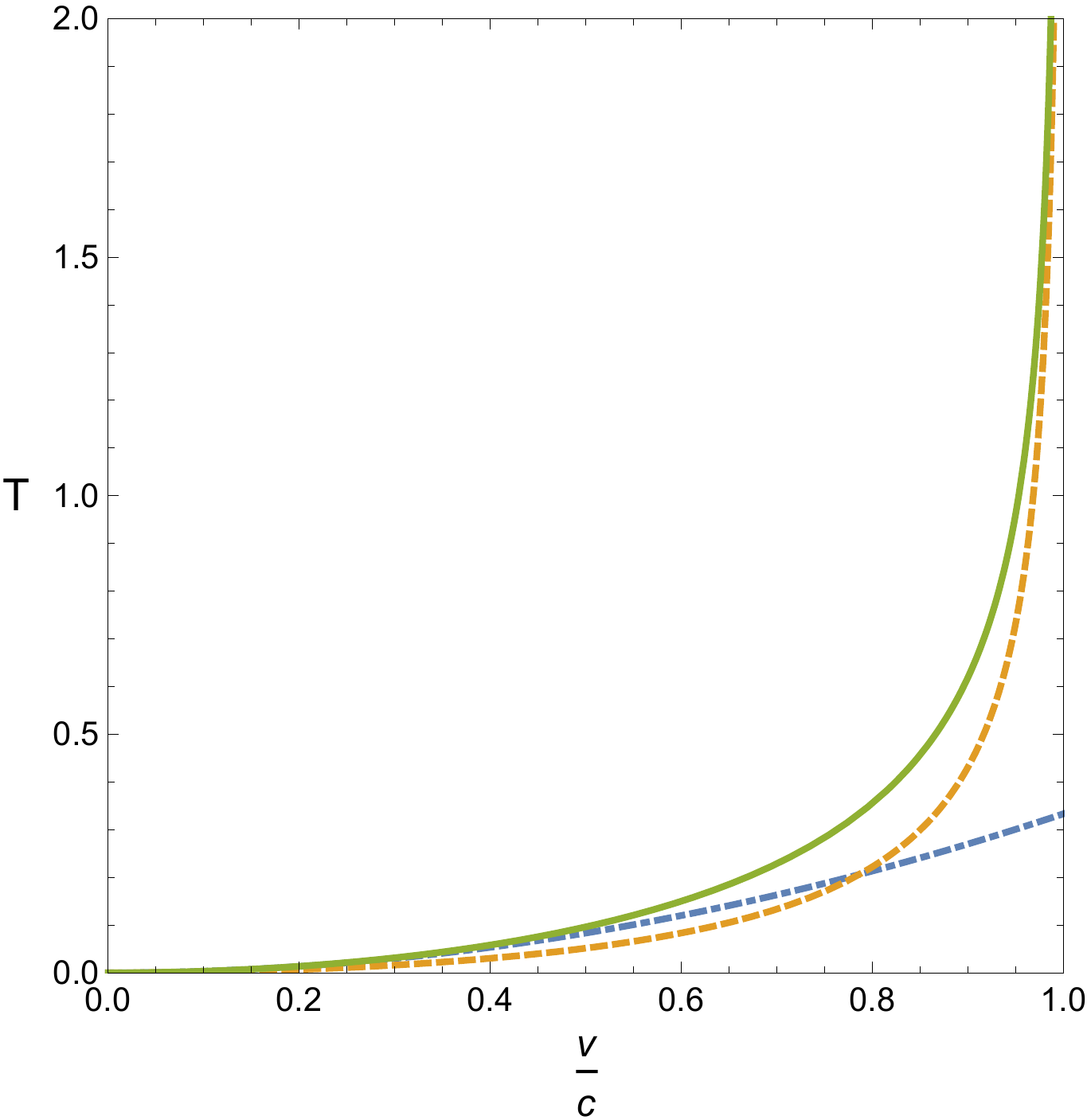}
\caption{\label{temperatures}Temperature as a function of average 
particle velocities. The green solid line 
is the true temperature calculated according to Eq. (\ref{temperature}).
	The blue dot-dashed line is the classical non-relativistic temperature of \mbox{Eq. (\ref{class_temp})}
	and the yellow dashed line is the
	temperature using the ultrarelativistic limit of Eq. (\ref{ultra_temp}).}
\end{figure}
Using $\mathcal{H} = \sum_{i=1}^N \sqrt{p_{i}^2 c^2 + m_i^2 c^4} + V$ this leads for 
every particle $i$ to 
\begin{equation}
\label{temperature}
	3 k_{\rm B} T = \langle \frac{p_i^2}{m \sqrt{1+\frac{p_i^2}{m_i^2 c^2}}}\rangle 
= m_i c^2 \langle \gamma_i - \frac{1}{\gamma_i} \rangle 
\end{equation}
with $p_i^2 = p_{i, x}^2 + p_{i, y}^2 + p_{i, z}^2$ meaning the square of the norm of the 3-dimensional 
momentum vector of particle $i$.
In the ultrarelativistic limit where $p \gg mc$ this yields the
relation
\begin{equation}
	\label{ultra_temp}
	k_{\rm B} T = \frac{1}{3N}\langle E_{kin}^{(rel)}\rangle
\end{equation}
where $E_{kin}^{(rel)} \approx m c^2 \gamma$.

As can be seen in fig. (\ref{temperatures}) the temperature
in special relativity deviates significantly from the classical ideal gas
law if the particles approach the speed of light.

\section{Molecular Dynamics simulation}
We simulate a system of $N$ particles interacting with 
a short-ranged soft potential in three dimensions.
The trajectories of
the particles evolve according to the relativistic equations of motion:
\begin{align}
	&\dot{\vec{q}}_{i} = \frac{\vec{p}_{i}}{m \sqrt{1 + \frac{p_i^2}{c^2 m_i^2}}} && \dot{\vec{p}}_{i} = \vec{F}_{i}
\end{align}
where $\vec{q}_{i}$ is the 3-dimensional position of particle $i$ and $\vec{p}_{i}$ the momentum respectively.

The force is then calculated from $\vec{F}_i = -\vec{\nabla}_i V$.
For $V$, the well-known Lennard-Jones potential is used, 
centered at each particle.
It reads in terms of the particle distance
\begin{equation}
\label{lennardjones}
	V(r) = 4 \epsilon \left(\left(\frac{\sigma}{r}\right)^{12} - \left(\frac{\sigma}{r}\right)^6 \right)
\end{equation}
where $r$ is the distance between particles and $\sigma$ and $\epsilon$
are parameters.

The accuracy of the method is checked by evaluating the energy

\begin{equation}
	\label{energy}
	\mathcal{H}=\sum_{i}\sqrt{p_i^2 c^2 + m_i^2 c^4} + V(q_1,\dots, q_n)
\end{equation}	
which needs to be conserved.
The potential is truncated at $r_{\text{cut}} = 2.5\sigma$. To avoid
discontinuities in the potential and its derivative (i.e. the force)
at $r_{\text{cut}}$, the potential is further modified by two terms:
\begin{equation}
\label{potential_cutoff}
	\tilde{V}(r)= V(r) - V(r_{\text{cut}}) - \left.\frac{\partial V}{\partial r}\right\rvert_{r_{\text{cut}}} \cdot (r-r_{\text{cut}})
\end{equation}
Since $V$ and its derivative are small quantities at $r_{\text{cut}}$ this does 
not change much the overall shape of the potential.
By employing the linked-cell method the complexity can be reduced
to $\mathcal{O}(N)$ \cite{linked_cell}.

\begin{figure}[h]
\centering
\includegraphics[width=0.45\textwidth]{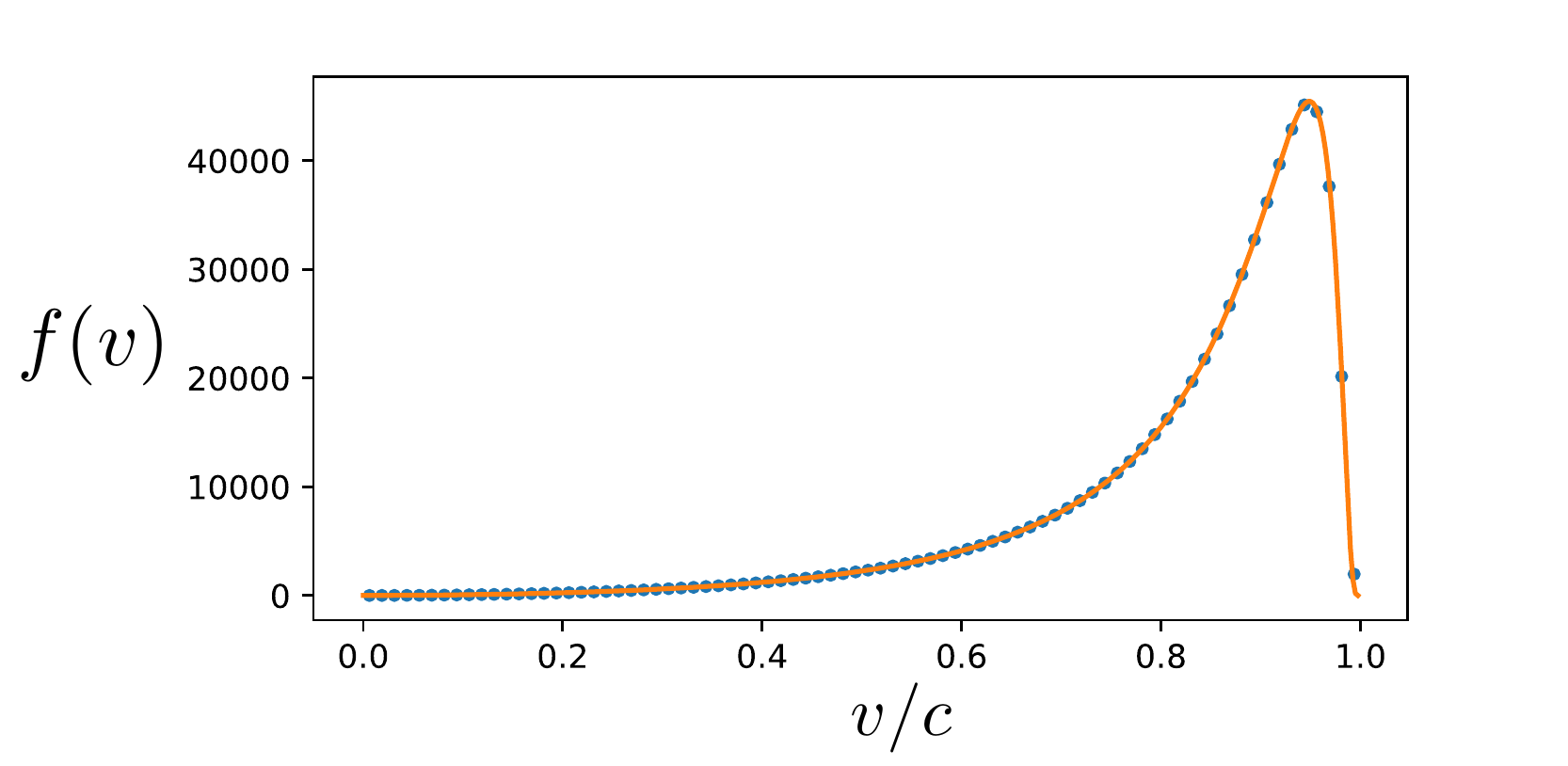}
\caption{\label{results1} Simulation results for the velocity distribution
of $N=8000$ particles. The horizontal axis
	is the velocity measured in units of the speed of light $c$. The blue
	points are a histogram obtained from the simulation data and the orange
	curve is the theoretical Jüttner distribution 
	with temperature $T=605 \epsilon / k_{\rm B}$ and $m=10^3 \epsilon / c^2$. 
	$\epsilon$ is the parameter from the potential (\ref{lennardjones}).
	}

\end{figure}

First, only the repelling part of the Lennard-Jones potential
(\ref{lennardjones}) proportional to $r^{-12}$ is considered. 
A histogram is constructed by dividing the range of absolute particle velocities 
into intervals of equal length. 
The temperature is calculated using Eq. (\ref{temperature}).
As can be seen in fig. \ref{results1} the obtained data fits the 
Jüttner distribution very well.

\begin{figure}[h]
	\centering
	\includegraphics[width=0.45\textwidth]{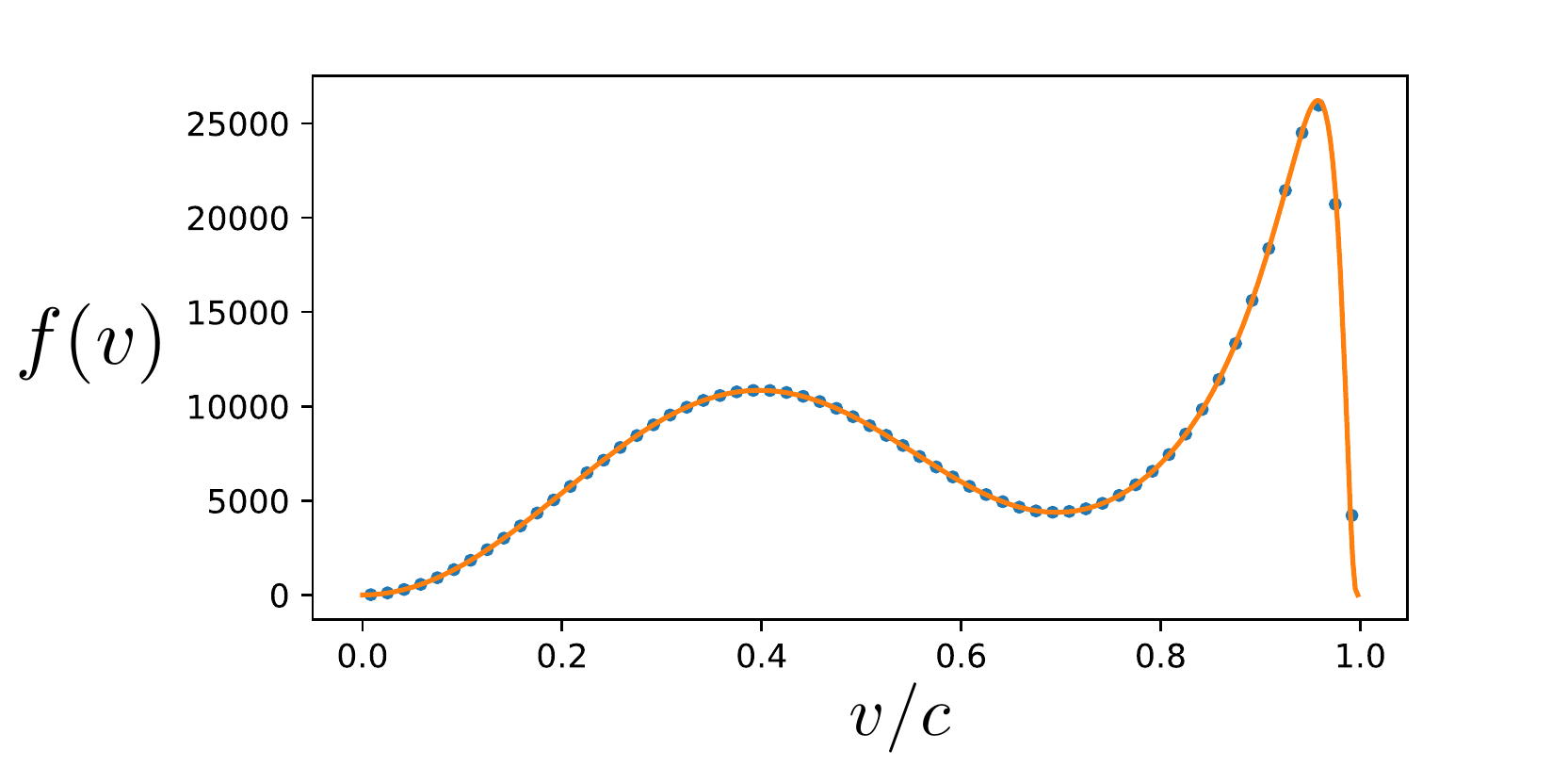}
	\caption{\label{twomasses} 
	Simulation with two sorts of particles differing in their masses.
	The parameters are: $N_1=N_2=4000$ particles with $m_1=10^4 \epsilon / c^2$ and $m_2=10^3 \epsilon / c^2$
	at temperature $T=668 \epsilon / k_{\rm B}$.
	The blue points are the histogram obtained from the simulation data
	and the orange line is the Jüttner distribution for two different masses \mbox{(Eq. (\ref{juttner_sum}))}.
	The heavy particles are responsible for the bulge at lower velocities.}
\end{figure}

A system of two different sorts of particles with different  
masses was simulated. In this case both sorts of particles should equilibrate
to the same temperature. The theoretical distribution is the sum of two Jüttner 
distributions for different masses but at the same temperature and normalised with
	the corresponding particle numbers $N_1$ and $N_2$:
	\begin{equation}
		\label{juttner_sum}
	f(v)=N_1 f_{m_1}(v)+N_2 f_{m_2}(v)
	\end{equation}
As can be seen in fig. \ref{twomasses} the particles are indeed fitted well
with the sum of two Jüttner distributions at the same temperature according to Eq. (\ref{juttner_sum}).

Also a system was considered where the masses of the particles
are randomly sampled from a Gaussian distribution $\rho$. In this
case the theoretical Jüttner distribution is a convolution
\begin{equation}
	\label{juttner_gauss}
	f(v)=\int \rho(m) f_m(v) dm
\end{equation}
with $f_m$ being the Jüttner distribution of particles with mass $m$.
And also in this case the velocity distribution 
is well-fitted by the Jüttner distribution of Eq. (\ref{juttner_gauss}).

\section{Relativistic thermostat}
Because the Hamiltonian equations of motion conserve the energy, the molecular 
dynamics simulations of the previous chapter simulate the microcanonical statistical
ensemble which is defined through constant total energy.
Such a system is closed in the sense that there is no energy exchange with an
environment. 
This situation is described by the microcanonical partition function
\begin{equation}
	\mathcal{Z}=\int \delta \left(\mathcal{H}(\vec{p}_1, \dots, \vec{q}_1, \dots) - E\right) d^3\vec{p}_1 \cdots d^3\vec{q}_1 \cdots 
\end{equation}
It is however more natural to allow the system to interact with its environment 
since experiments are normally conducted at constant temperature and not at constant 
energy \cite{nosehoover2}.
This situation is statistically described by the canonical ensemble. 
Here, the system is coupled to an infinite heat bath at a constant temperature $T$
which allows the system the exchange of energy.
The canonical partition function is:
\begin{equation}
	\mathcal{Z}=\int \exp\left(-\frac{\mathcal{H}(\vec{p}_1, \dots, \vec{q}_1, \dots)}{k_{\rm B} T}\right) d^3\vec{p}_1 \cdots d^3\vec{q}_1 \cdots
\end{equation}

The temperature of the system can be controlled using a thermostat.

Relativistic simulations with a stochastic thermostat have been conducted in Ref. \cite{montakhab}.
A thermostat that is both deterministic and samples from the canonical ensemble
is the Nosé-Hoover method \cite{nosehoover2}. Here 
a new degree of freedom $s$ is introduced that represents
the heat bath by adding a term proportional to $k_{\rm B} T \log(s)$ to the Hamiltonian. 
After applying Hamilton's equations of motion, a time transformation
needs to be done in order to arrive at the equations in real time
which makes the system non-Hamiltonian.
This gives rise to 
a friction parameter $\xi$ in the equations of motion
for the momenta.
\begin{equation}
	\label{p_friction}
\dot{p}=-\nabla V - \xi p
\end{equation}

Here we use the Nosé-Poincaré formalism proposed in \mbox{Ref. \cite{nosepoincare}}
in which the Hamiltonian does not require a time transformation
and the usual equations of motion \mbox{$\dot{\vec q}_i = \vec \nabla_{\vec p_i} \tilde{\mathcal{H}}$}
and \mbox{$\dot{\vec p}_i = -\vec \nabla_{\vec q_i} \tilde{\mathcal{H}}$} are directly valid.
According to this formalism a new Hamiltonian is introduced:
\begin{equation}
	\tilde{\mathcal{H}} = s \left( \mathcal{H}\left(\frac{\vec{p}_1}{s}, \dots,\vec{q}_1,\dots\right)
	+ \frac{p_s^2}{2Q} + 3N k_{\rm B} T \log(s) - \mathcal{H}_0\right)
\end{equation}
$\mathcal{H}$ is the original relativistic Hamiltonian (\mbox{Eq. \ref{energy}})
but as in the original Nosé-Hoover method the momenta are rescaled by
a factor of $s$.
The next two terms are responsible for the dynamics of the variable $s$ 
such that the canonical distribution is correctly sampled. $p_s$ is the 
conjugate momentum of the heat bath. 
The constant $\mathcal{H}_0$ is chosen such that ${\tilde{\mathcal{H}}}$ is zero 
at zero temperature.
Finally, the Hamiltonian is multiplied by $s$ such that the equations
of motion are obtained in real-time.

\begin{widetext}

	The Nosé-Poincaré thermostat applied to the relativistic Hamiltonian of Eq. \ref{energy} yields:
	\begin{equation}
		\tilde{\mathcal{H}} = s \left(\sum_i \sqrt{\frac{p_i^2}{s^2}c^2+m_i^2c^4} + V(\vec{q}_1, \dots) + 
		\frac{p_s^2}{2Q} + 3N k_{\rm B} T \log(s) - \mathcal{H}_0\right)
	\end{equation}
	We will now show that the microcanonical partition function of this extended Hamiltonian 
	is equivalent to the canonical partition function for the relativistic Hamiltonian:
	\begin{equation*}
		\mathcal{Z} = \frac{1}{N!}\int d^3 \vec{p}_1\cdots d^3 \vec{p}_n\int d^3\vec{r}_1\cdots d^3\vec{r}_n \int ds \int dp_s \delta(\tilde{\mathcal{H}}(\vec{p}_1,\dots,\vec{r}_1,\dots)-E)
	\end{equation*}
	Introducing the substitution $\tilde{\vec{p}}_i := \vec{p}_i / s$ and using the properties
	of the $\delta$-function yields:
	\begin{align*}
		\mathcal{Z} &= \frac{1}{N!}\int d^3\tilde{\vec{p}}_1\cdots d^3\tilde{\vec{p}}_n\int d^3\vec{r}_1\cdots \int ds \int dp_s
		s^{3N}\delta(\tilde{\mathcal{H}}(\tilde{\vec{p}}_1 \cdot s,\dots, \vec{r}_1,\dots) - E) \\ 
		&= \frac{1}{N!}\int d^3\tilde{\vec{p}}_1\cdots \int d^3\vec{r}_1\cdots \int ds \int dp_s s^{3N}
		\delta\left(s - \exp\left(-\frac{\sum_{n=1}^N \sqrt{\tilde{p}_i^2c^2 + m_i^2c^4} + V(r_1, \dots, r_n) + \frac{p_s^2}{2Q}}{k_{\rm B} T (3N)}\right)\right)\\
		&= \frac{1}{N!}\int d^3\tilde{\vec{p}}_1\cdots \int d^3\vec{r}_1\cdots 
		\exp\left(-\frac{\sum_{n=1}^N \sqrt {\tilde{p}_i^2 c^2 + m_i^2c^4} + V(\vec{r}_1, \dots, \vec{r}_n)}{k_{\rm B} T}\right) 
		\int dp_s
		\exp\left(-\frac{\frac{p_s^2}{2Q} - E}{k_{\rm B} T}\right)
	\end{align*}
\end{widetext}
Up to the constant factor from the last integral the canonical distribution function 
is obtained as we wanted to show.
The equations of motion are derived in the usual way through Hamilton's equations:
\begin{align*}
	\dot{\vec{r}}_i &=\frac{\vec{p}_i}{sm \sqrt{1+\frac{p^2}{s^2c^2m_i^2}}} \\
	\dot{\vec{p}}_i &= -s \vec{\nabla}_i V \\
	\dot{s} &=\frac{p_s}{Q} \\
	\dot{p_s} &= \sum_{i=1}^N \frac{p_i^2}{m s^2 \sqrt{1+\frac{p_i^2}{m^2c^2s^2}}} - 3N k_{\rm B} T(1+\log s) \\
	 &- V - \sum_{i=1}^N \sqrt{c^4m^2+ \frac{c^2p^2}{s^2}} - \frac{p_s^2}{2Q}+\mathcal{H}_0
\end{align*}

After reapplying the transformation $\tilde{\vec{p}}_i = \vec{p}_i / s$ they take the following
form:

\begin{align}
	\dot{\vec{q}}&=\frac{\tilde{\vec{p}}_i}{m \sqrt{1+\frac{\tilde{p}^2}{c^2m^2}}} \label{q_dot}\\
	\dot{\tilde{\vec{p}}}_i&=\vec{F}-\frac{\tilde{\vec{p}}_i}{s}\dot{s} \label{p_dot}\\
	\dot{s}&=\frac{p_s s}{Q} \label{s_dot}
\end{align}
\begin{align}
	\begin{split}
	\dot{p_s} &= \sum_{i=1}^N \frac{\tilde{p}_i^2}{m \sqrt{1+\frac{\tilde{p}_i^2}{m^2c^2}}} - 3N k_{\rm B} T(1+\log s) \\
		&- V - \sum_{i=1}^N \sqrt{c^4m^2+ c^2\tilde{p}^2} - \frac{p_s^2}{2Q}+\mathcal{H}_0 \label{ps_dot}
	\end{split}
\end{align}

Equation (\ref{q_dot}) is just the usual relativistic momentum-velocity relation.
Eq. (\ref{p_dot}) is of the form of Eq. (\ref{p_friction}).

\begin{figure}[h]
	\includegraphics[width=0.45\textwidth]{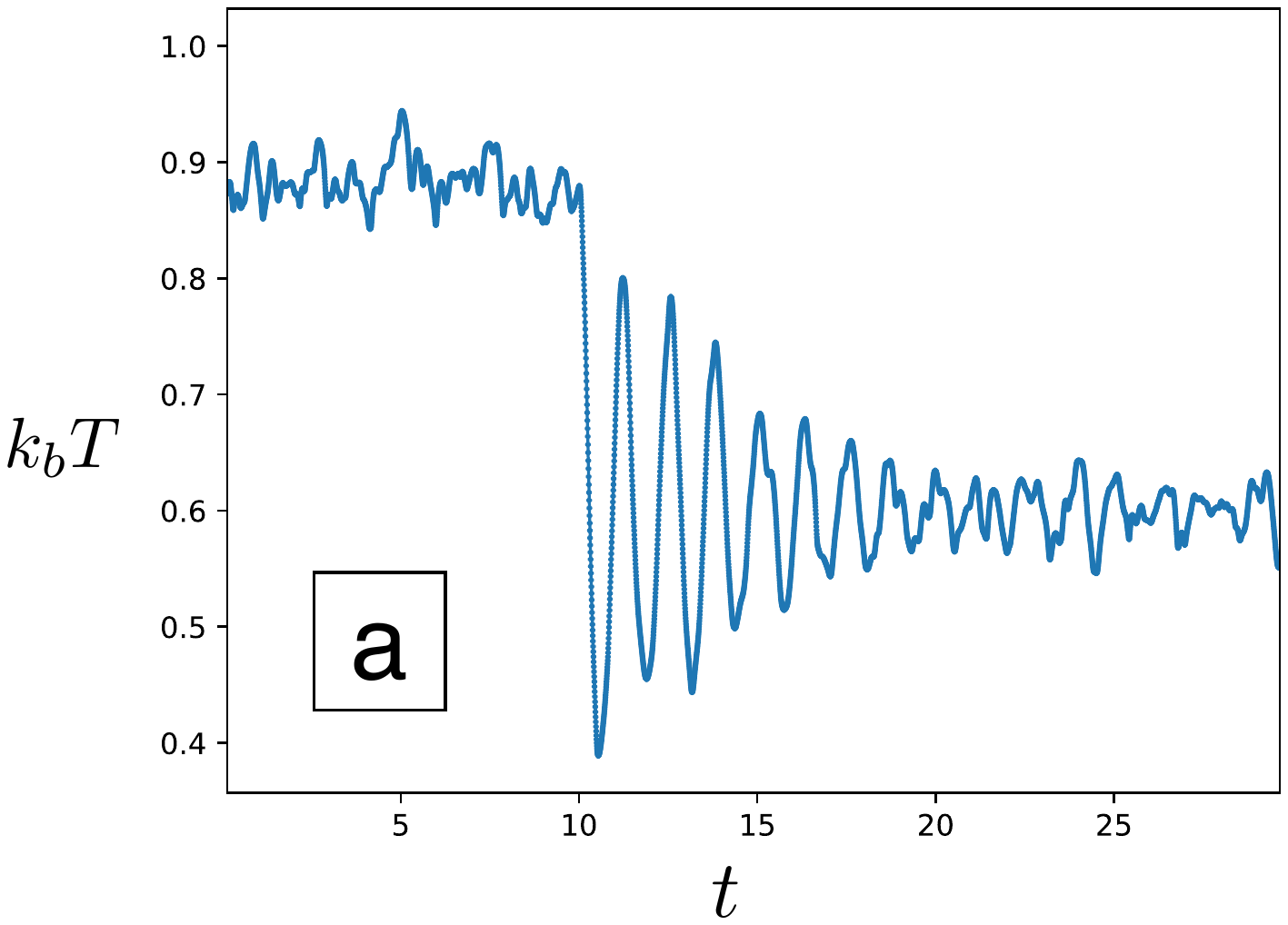}
	\includegraphics[width=0.46\textwidth]{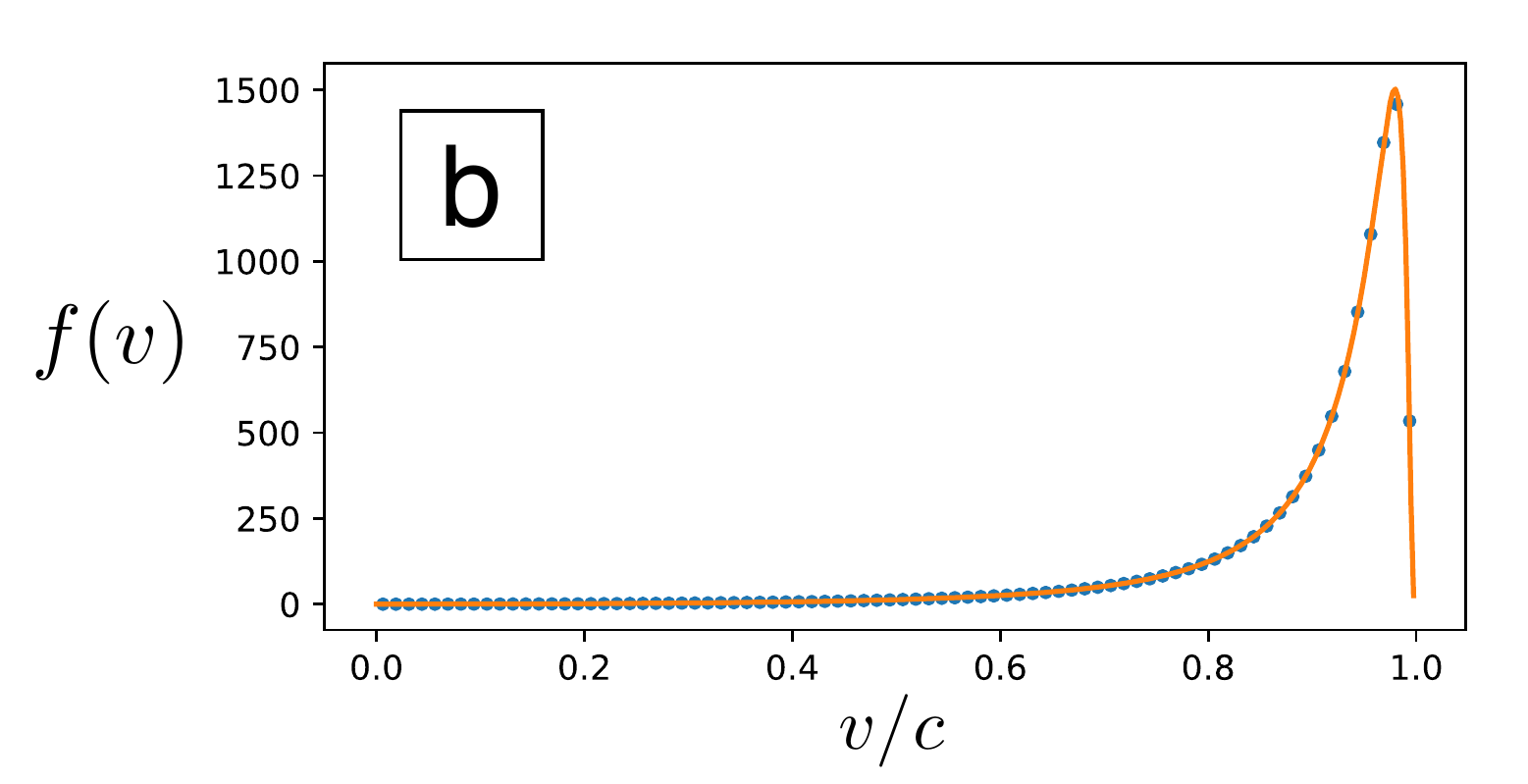}
	\caption{\label{temperature_thermostat}
	(a): Instantaneous temperature (measured in units $\epsilon / k_{\rm B}$) 
	over simulation time (for $N=125$ particles).
	A thermostat with temperature $T = 0.6 \epsilon / k_{\rm B}$ and $Q=10$ is applied 
	at $t = 10$ (measured in units $\sigma / c$). 
	(b): Jüttner distribution at the same temperature fitted
	to the simulation data obtained after the thermostat has relaxed ($t \gtrsim 20)$.}
\end{figure}

In Fig. \ref{temperature_thermostat}a the behaviour of the instantaneous temperature
(measured by evaluating \mbox{Eq. (\ref{temperature}))} of a system coupled to such a thermostat is shown. 
The simulation was conducted with
$N=125$ particles. Here we did not simulate larger systems, because we want
to study in the following statistical fluctuations.
The system first had temperature $T=0.89 \epsilon / k_{\rm B}$ and at time $10$
a different temperature $T = 0.6 \epsilon / k_{\rm B}$ was applied to the thermostat.
In Fig. \ref{temperature_thermostat}b we see that the correct Jüttner distribution
of the particle velocities is obtained. 

The behaviour of the thermostat depends on the choice of the thermal inertia $Q$. 
As can be seen from Eqs. (\ref{p_dot}), (\ref{s_dot}) and (\ref{ps_dot}),
the thermostat provides a feedback mechanism to control the 
temperature: If the temperature, represented by the first term of the right hand side in \mbox{Eq. (\ref{ps_dot})}
deviates from the chosen temperature, $p_s$ changes its value and thus following Eq. (\ref{p_dot})
slows down or speeds up the particles. If $Q$ is small the "friction" term in
Eq. (\ref{p_dot}) proportional to $\dot{s}/s = p_s$ changes fast and thus the feedback mechanism 
is very sensitive. On the other hand, if $Q$ is very large, the 
effect of the thermostat disappears.

\begin{figure}[h]
	\includegraphics[width=0.45\textwidth]{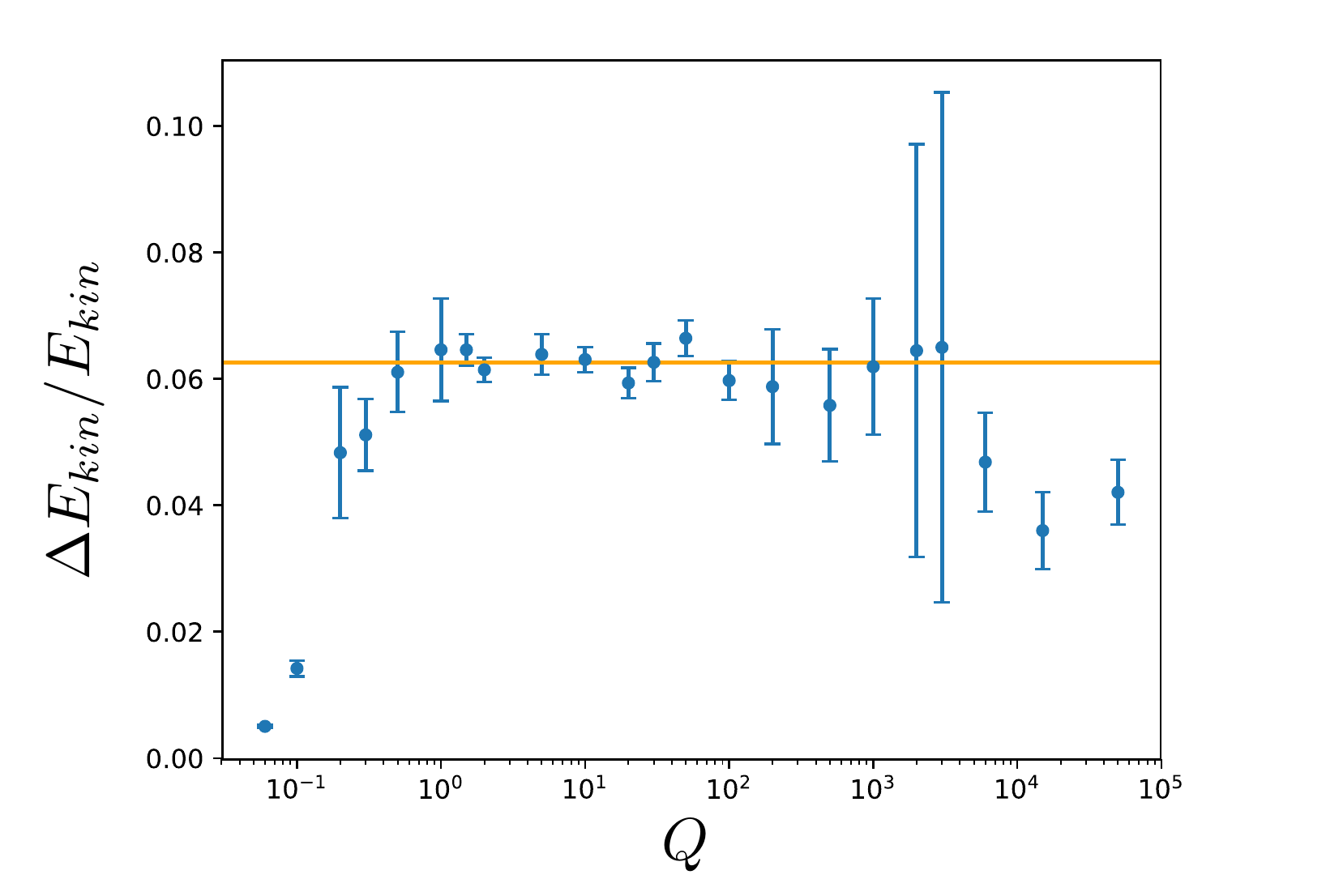}
	\caption{\label{fluctuations_energy}
	Kinetic energy fluctuations for different thermal inertias $Q$.
	$N=125$ particles, $T = 1 \epsilon / k_{\rm B}$, the orange line
	represents the theoretically predicted canonical energy fluctuations.
	The results are similar to those of Ref. \cite{fluctuations} where the fluctuations
	for the classical case were investigated.
	}
	
\end{figure}

A good choice
of $Q$ is characterised by the fact that canonical energy fluctuations
are obtained \cite{nosehoover}. The theoretical fluctuations in the canonical ensemble
are estimated by calculating $\Delta E_{kin} \approx \sqrt{Var(E_{kin})}$. 
The relative fluctuations  $\Delta E_{kin} / E_{kin}$ should be proportional 
to $1/\sqrt{N}$ where $N$ is the number of particles. The behaviour
of the kinetic energy fluctuations for a simulation with $N=125$ and $T = 1 \epsilon / k_{\rm B}$ 
is shown in fig. \ref{fluctuations_energy}.
The error bars represent the standard deviation.

In Ref. \cite{fluctuations} the effects of different $Q$ for a 
classical Nosé-Hoover thermostat were analyzed:
If $Q$ is set correctly,
the thermostat couples to the particle motion and the system
samples the canonical distribution.
However, if the value of $Q$ is too small or too large the 
variable $p_s$ oscillates periodically (with quick oscillations
if $Q$ is too small and slow oscillations if $Q$ is too large).
For very high $Q$ one actually simulates
the microcanonical ensemble \cite{fluctuations}.

For the thermostatted relativistic system a very similar result as in Ref. \cite{fluctuations}
is obtained. In a medium range  
of $Q$, the canonical fluctuations are reproduced (represented by the orange line in fig. \ref{fluctuations_energy});
if $Q$ exceeds this range,
the variance of the fluctuations increases and for very large $Q$ the fluctuations 
become smaller. If $Q$ is chosen very small, the kinetic energy fluctuations also become smaller,
however the kinetic energy oscillates with a high frequency.
In fig. \ref{fluctuations_energy} canonical energy fluctuations 
are observed for $1 < Q < 100$. 

For small values of $Q$ Nosé approximated the oscillations of the thermostat
with a harmonic oscillator \cite{nosehoover}. With a similar procedure
this can also be done for the relativistic thermostat:
Starting from Eqs. (\ref{p_dot}) and (\ref{s_dot}), it is assumed
that the motion of the particles is dominated by the thermostat. So we can neglect
the dependence of the force $F \approx 0$, so that all $p_i$ now have
the same equation of motion and thus we omit the index $i$ and $\alpha$ and write $p$.
This yields: 
\begin{equation}
	\label{p_dot_approx}
	\dot{p} \approx -p \frac{p_s}{Q}
\end{equation}
Eq. (\ref{ps_dot}) becomes:
\begin{equation}
	\label{ps_dot_approx}
	\dot{p_s} = \sum_{i=1}^N \frac{p^2}{m \sqrt{1+\frac{p^2}{m^2c^2}}} - 3 N k_{\rm B} T
\end{equation}
$p$ and $p_s$ are then linearly approximated around an equilibrium (time-independent)
solution $p = \langle p\rangle + \delta p$ and $p_s = \langle p_s\rangle + \delta p_s$.
The zero order solution is
\begin{align}
	\label{zero_order}
	3N \frac{\langle p\rangle ^2}{m \sqrt{1+\frac{\langle p\rangle ^2}{m^2c^2}}} &= 3 N k_{\rm B} T \\
	\langle p\rangle \cdot \langle p_s\rangle &= 0 \implies \langle p_s\rangle = 0
\end{align}
The first order solution yields:
\begin{align}
	\label{first_order}
	\delta \dot{p_s} &= 3N \left( \frac{2\langle p\rangle}{\sqrt{1+\frac{\langle p\rangle ^2}{m^2c^2}}} 
	- \frac{\langle p\rangle ^3}{m^3 c^2 (1+\frac{\langle p\rangle ^2}{m^2c^2})^{3/2}} \right) \delta p\\ \label{first_order2} 
	\delta \dot{p} &= -\frac{\langle p\rangle  \delta p_s}{Q} 
\end{align}
Taking the derivative of Eq. (\ref{first_order2}) and 
inserting it in Eq. (\ref{first_order}) yields (using Eq. (\ref{zero_order})):
\begin{equation}
	\label{oscillator}
	\delta \ddot{p} = -\underbrace{\frac{3N}{Q} \left( 2 T - \frac{T^2}{m c^2 \sqrt{1+ \frac{\langle p\rangle ^2}{m^2 c^2}}} \right)}_{(2\pi f)^2}\delta p
\end{equation}
which is the equation for a harmonic oscillator from
which the frequency $f$ can be inferred:
\begin{equation}
	\label{frequency}
	f = \frac{1}{2\pi}\sqrt{\frac{3N}{Q} \left( 2 T - \frac{T^2}{m c^2 \sqrt{1+ \frac{\langle p\rangle ^2}{m^2 c^2}}} \right)}
\end{equation}
Because of Eq.
(\ref{first_order2}) $\delta p_s$ oscillates with the same frequency.

\begin{center}
\begin{figure}[h]
\centering
	\includegraphics[width=.9\textwidth]{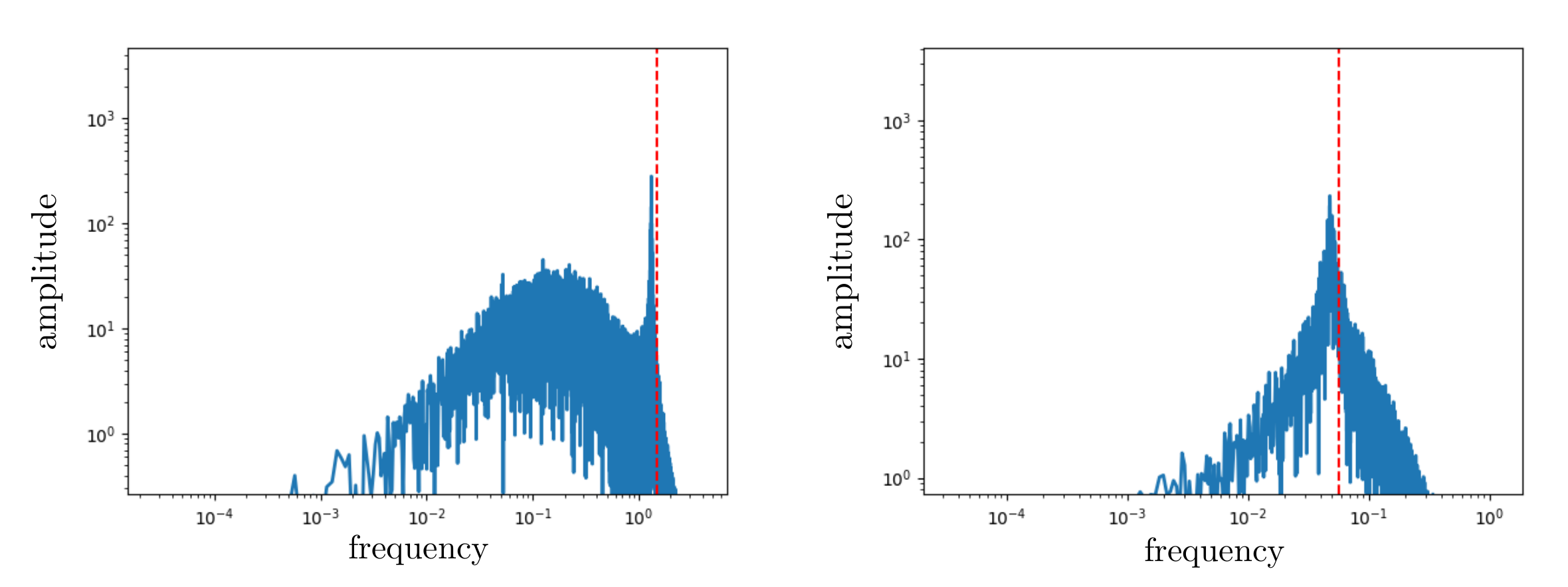}
	\caption{\label{fluctuations_analysis}
	Spectral analysis of the fluctuations of the variable $p_s$.
	The red line marks the oscillator frequency calculated from
	Eq. (\ref{frequency}). 
	Close below the red line, a pronounced peak can be found. 
	The other frequencies arise from interactions
	with the particle system. The simulation parameters are: $T = 1 \epsilon / k_{\rm B}$, $N=125$ and $m=1 \epsilon / c^2$. 
	The values for $Q$ are: $0.06$ (left) and
	$40$ (right).
	Canonical fluctuations are obtained
	if the oscillator frequency is in resonance with the particle system
	as on the left ($Q=40$).}
\end{figure}
\end{center}

The thermostat should be in resonance
with the system's fluctuations \cite{nosehoover2}. 
In Fig. (\ref{fluctuations_analysis}) a spectral analysis 
is shown of the time series
of $p_s$ for $Q = 0.06$ and $Q = 40$
with a relativistic temperature
$T = 1 \epsilon / k_{\rm B}$ and $m=1 \epsilon / c^2$.
The red line is the thermostat mode calculated from Eq. (\ref{frequency}).
There, the spectrum has a clear peak.
The other frequencies come from the particle system.
For small values of $Q$ the peak is sharper and
the thermostat is separated from the system. 
The thermostat mode should mix with the system's
modes and therefore a situation at the left of fig. \ref{fluctuations_analysis}
is desirable. In that case ($Q=40$), canonical energy fluctuations are observed
as can be seen in fig. \ref{fluctuations_energy}.
\onecolumngrid
\pagebreak
\twocolumngrid
\section{Conclusion}
Summarizing, the numerical experiments in three dimensions
confirmed that the Jüttner
distribution is a very good generalisation of the Maxwell-Boltzmann distribution
to special relativity. 
The results are independent
of the particle masses. 

The temperature was determined from the momenta of the particles using the equipartition theorem.
From the fact that we employed periodic boundary conditions 
arises the uniqueness of this definition: 
We have a special frame of reference (the boundaries) and the temperature
is determined from the momenta as measured in this frame.

Despite of the Jüttner function not being Lorentz invariant, it gives
the correct distribution in this frame of reference.
To investigate this further, 
one could generate histograms with constant rapidity bins as proposed in
Ref. \cite{evaldo} and compare them to a three dimensional 
generalisation of the Lorentz invariant distribution function that was derived in their paper.

The interaction between particles was assumed to happen instantaneously which
is theoretically not consistent with special relativity. 
In order to overcome this problem one could introduce fields and calculate
the force from a retarded potential. However, since the potential is short-ranged
it can be assumed that this effect 
will not result in a different velocity distribution.
Whether this has an effect on other quantities beside the velocity distribution, 
e.g. the heat capacity, could be investigated in further simulations.

A Nosé-Poincaré thermostat was coupled to a relativistic particle system. 
This allowed to control the temperature simulating the canonical ensemble.
The appropriate range for the thermal inertia $Q$ could be determined by monitoring the 
kinetic energy fluctuations. The behaviour of those fluctuations is similar to that of
a thermostatted classical system \cite{fluctuations} and in that regime they should
fulfill the fluctuation - dissipation theorem.
The thermostat itself exhibits a frequency that is visible 
in the spectrum of $p_s$ (which represents the friction parameter in the equations of motion). 
If $Q$ is in the appropriate range, the thermostat mode mixes with the frequencies 
of the particle system and thus the thermostat couples to the particles
and canonical energy fluctuations are obtained.
\bibliography{bericht6}
\end{document}